# Photoconduction in $Alq_3$

Debdutta Ray, Meghan P. Patankar, N. Periasamy and K. L. Narasimhan[a)].

*Tata Institute of Fundamental Research, Mumbai 400005, India.*

**Abstract**

Photoelectronic properties of $Alq_3$ were studied by photoconductivity (PC) measurements in thin film, sandwich (ITO/$Alq_3$/LiF/Al) devices. We find that the photocurrent is dominated by bulk generation of carriers for incident photon energies greater than 2.75 eV. The quantum efficiency of photocarrier generation has been measured from carrier collection measurements to be about 10%. The quantum efficiency is largely independent of electric field. This enables a direct measurement of the electric field dependence of mobility using photoconductivity measurements, which is used for quantitative analysis of the dark forward current in these devices. PC measurements were also used to obtain ($\mu_{0n}\,\tau_n$) product which can be used as a measure of material quality. For $Alq_3$, we find that the value of ($\mu_{0n}\,\tau_n$) product was between 3 x $10^{-15}$ $cm^2/V$ to 8 x $10^{-15}$ $cm^2/V$ for different samples. In forward bias, at high field the photocurrent shows saturation accompanied by a phase shift. These effects are attributed to space charge effects in the device.

[a)] Author to whom correspondence should be addressed; email: kln@tifr.res.in

## I. Introduction

The discovery of efficient electroluminescence in $Alq_3$ by Tang and Vanslyke[1] has revived interest in organic semiconductors and has given rise to many developments in the field of Organic Electronics. To be able to design efficient devices, it is necessary to understand transport and optoelectronic properties in organic semiconductors. These are low mobility semiconductors. Current is contact limited and transport is dominated by hopping of carriers from one molecule to another and the mechanism is poorly understood.[2] In organic semiconductors, photo-transport is even less well understood. An understanding of photo transport in organic semiconductors is also complicated by the fact that the optical excitations are generally excitonic.[2] To contribute to photocurrent, the excitons need to dissociate into charge transfer pairs and then free carriers. It is also important to decide whether the dissociation process takes place in the bulk or is catalysed at the electrode–semiconductor interface.[3,4] Although there has been a great deal of study related to transport in $Alq_3$,[4-10] there is very little literature on the photoelectronic properties of $Alq_3$.[11,12] Szmytkowski et al.[11] have measured the PC in $Alq_3$ and explain the PC as arising due to the quenching of excitons by the applied electric field. This assumes that the PC generation is a volume effect. Yang et al.[12] measure the spectral response of PC in $Alq_3$ and conclude from their measurements that the PC in $Alq_3$ can be explained by excitons diffusing to the electrodes followed by dissociation. We thus see that it is not clear whether the PC in $Alq_3$ is a volume effect or governed by the interface. This provided motivation for the work presented in this paper.

In this paper, we report on the photoconductivity in $Alq_3$. We show that the PC is a bulk effect and that a study of PC in $Alq_3$ enables us to separate out contact injection effects from transport process – both of which depend on the applied electric field and throws light on the transport process in $Alq_3$.

## II. Experiment

Thin films of $Alq_3$ were deposited in vacuum, at a pressure of 5 x $10^{-7}$ Torr, on patterned ITO coated glass substrates followed by a 10 Å LiF layer and 140 Å Al semitransparent cathode. The whole assembly was finally capped by a 1000 Å LiF layer. The LiF encapsulating layer was found to protect the Al electrode from oxidizing when working at large voltages. All the depositions were carried out in situ sequentially without exposing the samples to air. The samples were briefly exposed to air while transferring to a measurement jig. All measurements were carried out in a cryostat in vacuum at a pressure of $10^{-3}$ Torr. A 100 W xenon lamp coupled to a monochromator, a 632 nm He-Ne laser and a 400 nm laser diode (Nichia) were used as light sources. The light was mechanically chopped at 11 Hz. The photocurrent was measured using a lock in amplifier for the ac measurements and a Keithley electrometer for dc measurements. The photocurrent (PC) is defined as $J_{pc} = J_{light} - J_{dark}$. In the experiments reported here, the bias voltage was applied to the ITO electrode. The lamp response was measured using a thermopile detector. A calibrated silicon photodiode (Liconix, model 55PM) was used to measure the absolute intensity of the light falling on the sample. This was used in the estimation of quantum efficiency for the photocurrent. Unless otherwise mentioned, the experiments were done with the light incident through the semitransparent Al electrode. The percentage transmission through the Al electrode was around 60%, and data has been corrected accordingly. Experiments to measure the exciton dissociation efficiency under applied electric field were carried out using 400 nm light from a laser diode focused onto the sample electrodes. For these experiments, the sample area was 2mm x 2mm and the light focused down to 1mm diameter spot ensuring that all the light was transmitted through the sample. A fiber optic cable placed just behind the sample was coupled to a large area Si detector to measure the integrated photoluminescence (PL) signal. The PL quenching was measured as a function of applied electric field using square wave modulation. The modulated PL was detected using standard lock-in techniques. The measurements were done in reverse bias thereby avoiding interference from electroluminescence signals.

## III. Results and Discussion

In section A, we first present results to show that the PC is a bulk effect. We then present and discuss the PC in reverse bias and then in forward bias in sections B and C respectively. Finally, we use these results in section D to separate out injection effects from transport and analyze the dark forward bias current.

### *A. Origin of Photocurrent*

Figure 1 shows the modulated photocurrent as a function of applied bias for a 5000Å thick $Alq_3$ device at three different photon energies. At 2 eV (632 nm) excitation, the PC is asymmetric with bias and depends strongly on the sign of the applied bias. This is in contrast to excitation at 2.75 eV (450 nm) where the PC is symmetric with bias. Figure 1 also shows PC-Voltage data for excitation at 2.5 eV (500 nm), which shows an intermediate behavior. These results are understood as follows.

Optical excitation in organic semiconductors is excitonic in nature. To give rise to PC, the exciton has to be dissociated. Exciton dissociation can either take place in the bulk or at the electrode interface.[3,4,13,14] If exciton dissociation occurs at the interface, then excitons generated within a diffusion length from the electrode semiconductor interface can diffuse to the electrode, dissociate into free carriers and give rise to PC. The efficiency of exciton dissociation is dependent on the electrode. If exciton dissociation at the electrode is the only source of PC, then for asymmetric electrodes, the PC–Voltage characteristics would be asymmetric and strongly depend on the sign of the voltage bias.[14] On the other hand if the PC arises from a bulk process, then the PC density ($J_{pc}$) is expected to be given by,

$$J_{pc} = qG(X_n + X_p), \quad (1)$$

where G is the generation rate and X the collection length.[15,16] The collection lengths are $X_n = \mu_n \tau_n F$ and $X_p = \mu_p \tau_p F$ where F is the electric field, $\tau$ the carrier lifetime and $\mu$ the carrier mobility. The suffix n (p) stands for electrons (holes). Eq.1 is valid when the collection length is smaller than the sample thickness. Therefore for samples that absorb light uniformly, the PC should be independent of the sign of the bias voltage since $X_n$ and $X_p$ are the same for ±V.

From the above discussion and the data in fig.1, we conclude that the PC at 2 eV is dominated by exciton dissociation at the electrodes and at 2.75 eV by bulk photogeneration. We find that the Al electrode is more efficient than ITO at breaking up the exciton.

Figure 2 compares the PC for 3.25 eV (380 nm) light illuminated through the ITO anode and the Aluminum cathode respectively for a 5000Å thick $Alq_3$ device. The optical absorption coefficient at 380 nm is ~5 x $10^4$ $cm^{-1}$. In this case only 10% of the light reaches the back electrode. The PC is similar for both the cases. This result strengthens our conclusion that the measured PC in $Alq_3$ is a bulk process (for excitation energy greater than 2.75 eV).

Additional evidence that the photoconductivity is due to bulk carrier generation and not due to exciton dissociation at the electrode comes from a study of the spectral response. If exciton dissociation at the electrodes dominates the PC, then it is well known that only excitons generated within a diffusion length ($l_e$) will contribute to the PC. $l_e$ typically is about 10nm for many polymers and for $Alq_3$.[14,17] Hence only light absorbed in a distance $l_e$ will contribute to the PC and is independent of the sample thickness d, for $d > l_e$.[14] The spectral response of the sample will be independent of the sample thickness and symbatic with the absorption.

If, on the other hand, the PC arises due to a volume process, the spectral response of the PC would depend on sample thickness. Figure 3 shows the spectral response for samples of three different thickness viz. 500 Å, 1000 Å and 5000 Å respectively in reverse bias. Fig.3 also shows the optical absorption spectrum of $Alq_3$. We see from fig.3 that the PC action spectrum depends on the sample thickness. For example, for the 5000 Å sample, for E > 3.2 eV, 90% of the light is absorbed in the sample. We hence do not see any spectral features.

We now briefly comment on the PC excited with 2 eV (632 nm) light. We have presented evidence that the PC is due to excitons dissociating at the electrode interface. The bulk absorption coefficient ($\alpha_b$) of $Alq_3$ at 2 eV is very small and expected to be less than 1 $cm^{-1}$.[18] Hence, $\alpha_b\, l_e \leq 10^{-6}$. We estimate that the PC should be smaller by at least 2 orders than what is observed in fig. 1. An understanding of this discrepancy arises from the following. In thin films of $Alq_3$, the measured optical absorption coefficient below 2.5 eV is about 1000 times larger than $\alpha_b$ and is independent of energy below 2.5 eV, up to at least 1 eV.[18] The excess absorption has been understood as arising from surface states or a thin layer near the interface. This satisfactorily accounts for the larger than expected PC at 2eV. Hence the PC can be used to as a sensitive way of measuring the surface state absorption.

These experiments conclusively demonstrate that for E > 2.75 eV, the PC is due to bulk generation. The inset in fig. 3 shows that the photocurrent varies linearly with light intensity. We now focus on understanding the phototransport in $Alq_3$.

*B. Photocurrent in Reverse Bias*

In reverse bias, the maximum gain of the PC is limited to unity. Hence injection effects, which are important in forward bias, do not complicate the understanding of PC in reverse bias.[15,16] We see from eq.1, that the PC should be ohmic and proportional to the electric field. In organic semiconductors, the mobility ($\mu$) depends on the electric field and can be written as [9],

$$\mu = \mu_0 \exp(\beta\sqrt{F}) . \tag{2}$$

At low fields ($F < 10^6$ V/cm), the electron mobility is greater than the hole mobility in $Alq_3$ and electrons dominate the electrical transport.[19] We may hence ignore the contribution of holes to the photocurrent. The free carrier (volume) generation rate is,

$$G = \frac{\eta_{pc} I_0 (1 - e^{-\alpha d})}{L} = \frac{I_A \eta_{pc}}{L} \tag{3}$$

where $\alpha$ is the optical absorption coefficient, d the sample thickness, $I_0$ the incident light intensity (per unit area), $I_A$ the number of photons absorbed per unit area and $\eta_{pc}$ the quantum efficiency for free carrier production. The excitation length L takes the smaller of the two values - d or $1/\alpha$. We can then write for the PC density, $J_{pc}$ as

$$J_{pc} = \frac{q I_0 \eta_{pc} \mu_{0n} e^{\beta\sqrt{F}} \tau_n F (1 - e^{-\alpha d})}{L} , \tag{4}$$

where $\tau_n$ is the electron lifetime. Figure 4 is a semilog plot of ($J_{pc}$ / F) vs $F^{1/2}$. The electric field has been corrected for the built-in voltage which was measured as the voltage at which the PC changes sign.[20] Typically the built-in voltage for ITO/$Alq_3$/Al samples was around 0.8V. For thick samples, this correction is not as important as it is for thin samples. We see from fig.4, that the PC satisfies eq.4 up to the highest applied field viz. $1.8 \times 10^6$ V/cm. $\beta$ as obtained from the slope (using eq.4) has a value of $4.6 \times 10^{-3}$ $(cm/V)^{1/2}$. This value of $\beta$ is consistent with that reported for electrons in the literature, obtained from transient electroluminescence measurements.[19] We hence conclude that the increase in $J_{pc}$ with bias is a consequence of the increase in the electron mobility with electric field. This suggests that $\eta_{pc}$ is independent of the electric field. The intercept in fig.4 is given by ($q\ I_A\ \eta_{pc}\ \mu_{0n}\ \tau_n / L$). Since L is known and $I_A$ can be independently measured, ($\eta_{pc}\ \mu_{0n}\ \tau_n$) can be easily calculated for these samples. This quantity is closely related to the figure of merit for a photoconductor ($\eta_{pc}\ \mu\ \tau_n$) and can be used as a measure of material quality.

In the case of 2 eV (632 nm) excitation, the PC arises due to exciton dissociation at the electrodes. A plot of log ($J_{pc}/F$) vs $F^{1/2}$ allows us to estimate $\beta$ for both forward and reverse bias for 2 eV light. The value of $\beta$ is similar to that obtained for 2.75 eV (450 nm) and 3.26 eV (380 nm) photon energies. We hence conclude that electrons are responsible for transport and the asymmetry in the magnitude of PC (for 2 eV excitation) is due to the difference in the quantum efficiency of exciton dissociation at the two interfaces.

In Figure 5 we show a plot of "normalized" photocurrent, ( $J_{pc} / ( I_0\ F \exp(\beta F^{1/2}) )$ ), as a function of photon energy for different applied electric fields. We see from the figure that the "normalized" photocurrent spectra are independent of the electric field. The optical absorption coefficient of $Alq_3$ is approximately constant ($5 \times 10^4$ $cm^{-1}$) between 3 eV to 3.6 eV. For a 5000 Å thick film, 90% of the light is absorbed in this energy range. The "normalized" photocurrent in figure 5 is independent of both electric field and photon energy in this energy range. It follows from eq.3 that the quantity ($\eta_{pc}\ \mu_{0n}\ \tau_n / L$) is independent of the electric field. We hence conclude that PC measurements allow us to separate out carrier injection effects from transport parameters in a conductivity measurement. In section C, we show that this can be used for quantitative analysis of the dark forward current. We now obtain the value of $\eta_{pc}$ from collection efficiency measurements.

When the collection length ($X_n + X_p$) is $\geq$ d, the sample thickness, for uniformly absorbed light the PC density ($J_{pc}$) is given by,

$$J_{pc} = Gqd\,, \tag{5}$$

and is independent of the mobility and the electric field.[15,16] This provides a direct estimation of $\eta_{pc}$ (eq. 3) using thin samples. Figure 6 shows the PC as a function of applied bias for a 500 Å device. The incident photon energy is 3.25 eV (380 nm). For F > 2 x $10^6$ V/cm, the PC saturates - implying that all the carriers are collected. From the saturated PC in reverse bias, we estimate $\eta_{pc}$ to be 10%. The $\mu_{0n}\tau_n$ product for all the films, of different thicknesses (500Å, 1000 Å and 5000 Å), was found to be in the range between 3 x $10^{-15}$ to 8 x $10^{-15}$ $cm^2/V$. The saturation of photocurrent implies that the carrier generation efficiency is not a strong function of electric field. This is confirmed from the results of photoluminescence quenching experiment reported later in this section.

All the above experiments were carried out with the device under reverse bias. The reverse bias PC of an insulator with non-injecting contacts has been understood in the framework of a simple model due to Goodman and Rose[15] and Crandall.[16] In this model, for uniformly absorbed light, the sample can be divided into three regions $X_n$, $X_p$ and $X_g$. This is shown schematically in figure 7. $X_n$ and $X_p$ are adjacent to the contacts. Recombination is negligible in these two regions and the current is governed by drift in these regions. In the region $X_g$, generation is equal to recombination. With increasing bias, $X_n$ and $X_p$ increase with a concomitant decrease of $X_g$. The net current density is hence given by G q ($X_p + X_n$). The collection length $X_n$ ($X_n = \mu_{0n} \exp(\beta F^{1/2})\, \tau_n F$) can be estimated using the $\mu_{0n}\tau_n$ product and $\beta$, obtained from our experimental data. $X_n$ = 40 Å and 550 Å for F = $10^6$ V/cm and 2 x $10^6$ V/cm respectively. At fields close to 2 x $10^6$ V/cm, the hole collection length also becomes appreciable and comparable to the electron collection length. This can be estimated from the drift mobilities of electrons and holes.[19] The data in figure 6 is consistent with the above description. Thus, $Alq_3$ behaves like a classical photoconductor.

In the above experiments, we have presented evidence that $\eta_{pc}\mu_{0n}\tau_n$ does not depend on the electric field. However, it is possible that $\eta_{pc}$ can be electric field dependent due to the dissociation of excitons in an applied electric field. This can be verified from photoluminescence quenching experiments. The photoluminescence (PL) quenching efficiency $\eta_q(F)$ is a direct measure of the increase in the photogeneration quantum efficiency arising from the dissociation of excitons due to the applied electric field. The PL quenching efficiency can be written as,

$$\eta_q(F) = \frac{PL(0) - PL(F)}{PL(0)}, \tag{6}$$

where PL(F) and PL(0) are the PL intensities in the presence and absence of an external field respectively.[11]

Figure 8 shows $\eta_q(F)$ as a function of electric field for a 5000Å thick sample in reverse bias excited at 3.1 eV (400 nm). We see from the figure that for F = 1.8 x $10^6$ V/cm, $\eta_q(F)$ ~ 1% in reverse bias, which is a very small fraction of $\eta_{pc}$ ~ 10%. Strictly speaking $\eta_{pc}$ is a sum of two terms, $\eta_0$ and $\eta_q(F)$, where $\eta_0$ is electric field independent which has a value of 9%. Thus,

$$\eta_{pc} = (\eta_0 + \eta_q(F)), \tag{7}$$

The variation of $\eta_0$ with photon energy is reported and discussed below.

Figure 9 is a plot of $\eta_0$ vs. the photon energy for the 5000 Å sample. This plot was obtained by measuring the photocurrent per absorbed photon at a fixed bias (50 V; F = $10^6$ V/cm). At this field $\eta_q(F)$ is negligible compared to $\eta_0$ and thus $\eta_{pc}$ ~ $\eta_0$. The absolute value of $\eta_0$ was obtained using the value of $\eta_0$ (9%) determined from the collection efficiency measurements (fig.6) at 3.26 eV (380 nm). The absorbance of the sample at low excitation energy was measured both from an optical absorption measurement and from a luminescence excitation measurement. The latter is more reliable when the absorption coefficient is small as it is immune to scattering.[17,18] Both measurements give similar results. $\eta_0$ vs. photon energy plot shows a peak at 2.75 eV (450 nm) which is 0.5 eV less than the optical absorption peak.

For molecular crystals, the yield $\eta_{pc}$ for bulk photoionisation is seen to satisfy a relation of the form,

$$\eta_{pc} \propto (h\nu - h\nu_0)^2, \tag{8}$$

where $h\nu_0$ is the photoionisation threshold energy.[14] The inset in fig.9 is a plot of $(\eta_{pc})^{1/2}$ vs photon energy. We see that for photon energy < 2.65 eV, the data satisfies eq.8 with $h\nu_0$ = 2.4 eV. That is, the threshold photon energy for the bulk photogeneration process is 2.4 eV. This is consistent with the data in fig.1, where we experimentally show that at 2.5 eV, the photoconduction has both bulk and interface contribution to PC.

We now briefly comment on the mechanism of PC in $Alq_3$. We note the following important results from our experiments:

1) The threshold for PC due to a bulk process in $Alq_3$ is at 2.4 eV.

2) The PC at photon energies < 2.4 eV is due to interface effects.

3) The quantum efficiency for bulk photocarrier generation ($\eta_{pc}$) has major contribution from the electric field independent term $\eta_0$.

We can interpret the photoconduction in $Alq_3$ as due to direct excitation of C-T pairs from the ground state. This C-T transition is weak and cannot be seen in an optical absorption experiment but is readily discernible in a photoconductivity measurement. Such transitions have been postulated for other systems.[21-23]

It is known that there are many isomers of $Alq_3$ present in thin films.[24-28] These isomers have different optical properties. The percentage of these isomers present in the film cannot be controlled easily. From TOF measurements, Malliaras et al.[24] has estimated the fac isomer to be about 12% in his samples. It is possible that the PC is dominated by one of these isomers which are present in a small but varying quantity in different films. Alternatively, one of these isomers could act as a bulk sensitizing agent to explain the PC results presented here. If this is indeed the case, then PC measurements as described in this paper could turn out to be a good way of quantifying the concentration of these isomers in $Alq_3$ based devices.

*C. Photocurrent in Forward Bias*

The difference between forward and reverse bias in a PC experiment is that it is possible to have photoconductive gain > 1 under forward bias. This is one source of asymmetry in the Photocurrent – Voltage characteristics between forward and reverse bias. This asymmetry is easily observed in thin samples (e.g. 500 Å) and not in thick samples. Figure 10 shows the PC-Voltage and Electroluminescence-Voltage characteristics for a 5000 Å thick $Alq_3$ device. A major difference in the Photocurrent – Voltage characteristics between forward and reverse bias is that under forward bias condition, we observe the saturation of the PC beyond 75 V (1.5 x $10^6$ V/cm). However, this saturation should not be mistaken for total charge collection as seen in the 500 Å sample in reverse bias (fig. 6). The saturation in forward bias, seen in fig. 10, is also seen in 1000 Å device at similar fields. The onset of the saturation is seen to be closely related to the onset of electroluminescence (fig.10) and is attributed to space charge effects discussed below.

Another interesting observation of the PC under forward bias relates to the phase of the PC signal. The photocurrent is expected to change sign when the applied field is equal in magnitude and opposite in sign to the built-in electric field ($V_{bi}$ = 0.8V).[20] This is seen in figure 11 which shows the phase of the PC as a function of applied bias for a 5000 Å thick $Alq_3$ device at room temperature. As we go to larger forward bias, we expect the phase of the PC signal to be independent of the magnitude of bias. In fig.11, we see (in forward bias) that the phase of the PC is constant up to 55V (F = 1.1 x $10^6$ V/cm). Beyond 55V, it exhibits a large phase shift. The onset of this phase shift is found to be closely related to the onset of electroluminescence, which takes place at a similar bias (fig. 10).

We attribute both the above observations (onset of PC saturation in forward bias and phase shift) to space charge effects in the sample. This is explained as follows. In forward bias, in the dark electrons are injected from the cathode. The field in the sample is modified by the presence of the electron space charge. At larger forward bias, holes are injected at the anode. This modifies the space charge and decreases the field in the vicinity of the contacts thereby influencing the magnitude of $X_n$ and $X_p$. We suggest that the saturation in the PC in forward bias is related to the influence of the space charge. This can also be responsible for the phase shift seen in forward bias. Detailed simulations to calculate the effect of space charges is necessary to reach a satisfactory conclusion.

*D. Dark Current in Forward Bias*

We have seen in Section B that the PC measurements under reverse bias help us to separate carrier transport from carrier generation. In this section, we will use these results to clarify our understanding of the dark current under forward bias. Specifically, we will attempt the separation of contact injection effects from transport.

In forward bias, the dark current density $J_d$ can be written as

$$J_d = qn\mu F, \tag{9}$$

where n is the carrier concentration. The equilibrium carrier concentration in $Alq_3$ is very small. Under forward bias conditions, electrons are injected from the cathode into $Alq_3$. The carrier concentration is a field dependent quantity as the cathode is not an ohmic but a Schottky contact. Carrier injection takes place as a consequence of image force lowering of the Schottky barrier. The barrier lowering $\delta\varphi$ can be written as $\delta\varphi = (qF/4\pi\varepsilon\varepsilon_0)^{1/2}$.[29] If the current is limited only by injection from the cathode, then assuming a thermionic emission model to describe the injection over the barrier and including image force lowering terms, the forward current $J_d$ can be written as,

$$J_d = AT^2 \exp\left(-\frac{q\phi_b - \beta_{RS}\sqrt{F}}{kT}\right), \tag{10}$$

where A is the Richardson constant, $(q\ \varphi_b)$ the barrier height and $\beta_{RS} = (q^3 / 4\pi\varepsilon\varepsilon_0)^{1/2}$.[29] At room temperature, $\beta_{RS}/kT$ can be calculated to have a value of $8.5 \times 10^{-3}$ $(cm/V)^{1/2}$ for $Alq_3$ using $\varepsilon = 3.5$.[4] For $Alq_3$, a plot of log $J_d$ vs $F^{1/2}$ is a straight line with a slope of $1.3 \times 10^{-2}$ $(cm/V)^{1/2}$. Similar values have been reported in the literature.[6,7] Since the experimental value ($1.3 \times 10^{-2}$ $(cm/V)^{1/2}$) is 1.5 times larger than the theoretical value for $\beta_{RS}/kT$, it was argued that the forward current cannot be entirely described by the simple Schottky barrier theory.[6] We agree with this argument and now give a quantitative analysis of $J_d$ that gives a correct value for $\beta_{RS}/kT$.

In the above analysis, it was assumed that the current was determined only by the carrier injection efficiency. However, if both carrier injection and transport are important in the determination of the dark forward current, then we write as before for the density of injected carriers, n, as,

$$n = C\exp\left(-\frac{q\phi_b - \beta_{RS}\sqrt{F}}{kT}\right), \tag{11}$$

where we have taken into account the image force lowering terms. C is constant in electric field. Thus from equations 2, 9 and 11, we can write for the forward dark current, $J_d$, as,

$$J_d = qC\mu_0 F \exp\left(\beta\sqrt{F} + \frac{\beta_{RS}\sqrt{F} - q\phi_b}{KT}\right), \tag{12}$$

A plot of log ($J_d$ / F) vs $F^{1/2}$ is expected to be a straight line having a slope of (($\beta_{RS}$ / kT) + $\beta$). Fig. 12 shows the semilog plot of ($J_d$ / F) vs $F^{1/2}$. The value of (($\beta_{RS}$ / kT) + $\beta$) is found to be 1.31 x $10^{-2}$ $(cm/V)^{1/2}$. Using the value of $\beta$ obtained from PC measurements (4.6 x $10^{-3}$ $(cm/V)^{1/2}$), we find $\beta_{RS}$ / kT to be 8.5 x $10^{-3}$ $(cm/V)^{1/2}$, which is in excellent agreement with the theoretical value. We hence conclude that image force lowering satisfactorily accounts for the injection of carriers from the cathode in $Alq_3$ if we include the field dependence of the electron mobility.

## IV. Conclusion

In conclusion, we have measured the photoconductivity in thin films of $Alq_3$ in sandwich geometry as a function of excitation energy and externally applied electric field. We find that the photogeneration in $Alq_3$ is a bulk process and is not dominated by exciton dissociation at the electrodes for excitation energy above 2.75 eV. The field dependence of the photocurrent can be satisfactorily accounted for by the field dependence of the electron carrier mobility. PC experiments can hence be used to get information about the carrier mobility available hitherto only from TOF measurements. We show that $Alq_3$ behaves like a classical photoconductor - that is, in reverse bias, it is possible to measure a saturated photocurrent corresponding to total carrier collection. In reverse bias, PC can be used to obtain the ($\eta$ $\mu_{0n}$ $\tau_n$) product. This is a measure of material quality and can be used for quantitative comparison of different samples. The photocurrent under forward bias saturates beyond an electric field of 1.5 x $10^6$ V/cm. This is accompanied by a phase shift. These effects are attributed to space charge effects in the sample. Using PC data we show that it is possible to separate out carrier injection effects from transport in forward bias.

## Reference

1. C. W. Tang and S. A. VanSlyke, Appl. Phys. Lett. **51**, 913 (1987).
2. Martin Pope and Charles E. Swenberg, *Electronic Processes in Organic Crystals and Polymers*, 2nd ed. (Oxford University Press, New York, 1999), pp. 273-455.
3. Z.D. Popovic, A. M. Hor, and R. O. Loutfy, Chem. Phys. **127**, 451 (1988).
4. S. Barth, H. Bassler, T. Wehrmeister and K. Mullen, J. Chem. Phys. **106**, 321 (1997).
5. A. Ioannidis, E. Forsythe, Yongli Gao, M. W. Wu, E. M. Conwell, Appl. Phys. Lett. **72**, 3038 (1998).
6. U. Wolf, S. Barth, and H. Bassler, Appl. Phys. Lett. **75**, 2035 (1999).
7. Anver Aziz and K. L. Narasimhan, J. Appl. Phys. **88**, 4739 (2000).
8. P. E. Burrows, Z. Shen, V. Bulovic, D. M. McCarty, S. R. Forrest, J. A. Cronin, and M. E. Thompson, J. Appl. Phys. **79**, 7991 (1996).
9. Wolfgang Brutting, Stefan Berleb, and Anton G. Muckl, Organic Electronics **2**, 1 (2001).
10. W. Brutting, S. Berleb, A. G. Muckl, Synth. Met. **122**, 99 (2001).
11. J. Szmytkowski, W. Stampor, J. Kalinowski, and Z. H. Kafafi, Appl. Phys. Lett. **80**, 1465 (2002).
12. C. L. Yang, Z. K. Tang, W. K. Ge, J. N. Wang, Z. L. Zhang, and X. Y. Jian, Appl. Phys. Lett. **83**, 1737 (2003).
13. K. C. Kao and W. Hwang, *Electrical Transport in Solids* (Pergamon Press Inc., Oxford, 1981).
14. S. Barth, H. Bassler, H. Rost, and H. H. Horhold, Phys. Rev. B **56**, 3844 (1997).
15. Alvin M. Goodman and Albert Rose, J. Appl. Phys. **42**, 2823 (1971).
16. Richard S. Crandall, in *Semiconductors and Semimetals*, Vol. 21, Part B, edited by Jacques I. Pankove (Academic Press, Inc., Orlando, 1984), pp. 245-297.
17. D. Z. Garbuzov, V. Bulovic, P. E. Burrows, and S. R. Forrest, Chem. Phys. Lett. **249**, 433 (1996).
18. Anver Aziz and K. L. Narasimhan, Synth. Met. **122**, 53 (2001).
19. Anton G. Muckl, Stefan Berleb, Wolfgang Brutting, and Markus Schwoerer, Synth. Met. **111-112**, 91 (2000).
20. Debdutta Ray, Meghan P. Patankar, N. Periasamy and K. L. Narasimhan, unpublished.
21. L. Sebastian, G. Weiser, G. Peter, and H. Bassler, Chem. Phys. **75**, 103 (1983).

22. P. J. Bounds and W. Siebrand, Chem. Phys. Lett. **75**, 414 (1980).

23. M. G. Harrison, J. Gruner, and G. C. W. Spencer, Phys. Rev. B **55**, 7831 (1997).

24. George G. Malliaras, Yulong Shen, David H. Dunlap, Hideyuki Murata, and Zakya H. Kafafi, Appl. Phys. Lett. **79**, 2582 (2001).

25. M. Brinkmann, G. Gadret, M. Muccini, C. Taliani, N. Masciocchi, and A. Sironi, J. Am. Chem. Soc. **122**, 5147 (2002).

26. Michele Muccini, Maria Antonietta Loi, Kester Kenevey, Roberto Zamboni , Norberto Masciocchi, and Angelo Sironi, Adv. Mater. **16**, 861 (2004).

27. M. Braun, J. Chem. Phys. **114**, 9625 (2001).

28. Michael Colle, Robbert E. Dinnebier, and Wolfgang Brutting, Chem Comm, 2908 (2002).

29. S.M. Sze, *Physics of Semiconductor Devices*, 2nd ed. (Wiley-Interscience, New York, 1981), pp. 245-265.

**Figure captions**

**Fig. 1.** The photocurrent density ($J_{pc}$) is shown as a function of voltage, for a 5000 Å thick $Alq_3$ device, when illuminated by light of photon energies 2eV (632 nm) (△), 2.5eV (500 nm) (□) and 2.75eV (450nm) (○).

**Fig. 2.** A plot of Photocurrent density ($J_{pc}$) vs Voltage for 3.25 eV (380 nm) light incident through Al (△) and ITO (○) electrodes for a 5000 A thick $Alq_3$ device.

**Fig. 3.** A plot of photocurrent action spectra, in reverse bias, for $Alq_3$ devices of thickness 500 Å (□), 1000 Å (▲) and 5000 Å (○). The optical absorption spectrum (solid line) of the sample is also shown for comparison. The inset shows the plot of photocurrent as a function of incident light intensity.

**Fig. 4.** A semilog plot of ($J_{pc}$ / F) vs $F^{1/2}$ for a 5000 Å thick $Alq_3$ device in reverse bias. The photon energy for excitation was 3.25 eV (380 nm).

**Fig. 5.** Plot of $J_{pc} / I_0 F \exp(\beta F^{1/2})$ vs. incident photon energy for different electric fields across a 5000 Å thick $Alq_3$ device; 1) $2 \times 10^5$ V/cm (□), 2) $6 \times 10^5$ V/cm (△), 3) $1 \times 10^6$ V/cm (○) and 4) $1.8 \times 10^6$ V/cm (▽). The inset is the plot of the data on a semi-log scale. All the plots collapse on top of each other, suggesting that the carrier generation efficiency is independent of the electric field.

**Fig. 6.** A plot of photocurrent density ($J_{pc}$) as a function of voltage in reverse bias for a 500 Å thick (thin) $Alq_3$ device. The device was illuminated using 3.25 eV (380 nm) light. The saturation is due to collection of all photo-generated carriers (see text).

**Fig. 7.** Spatial map of photocurrent densities for an insulator with non-injecting contacts and uniform photon absorption (from ref. 15).

**Fig. 8.** Plot of Electric field dependent exciton quenching efficiency ($\eta_q$ (F)) vs applied electric field for a 5000 Å thick $Alq_3$ device. The incident photon energy was 3.1 eV (400 nm). The experiment was done in reverse bias.

**Fig. 9.** A plot of photocurrent quantum efficiency ($\eta_0$) vs photon energy as obtained from spectral response of photocurrent and number of photons absorbed, for a 5000 Å thick $Alq_3$ device. The inset shows the plot of $\eta_0^{1/2}$ vs photon energy near the photocurrent threshold ($<$ 2.65 eV).

**Fig. 10.** Plot of the photocurrent density ($J_{pc}$) (○) and electroluminescence (solid line) as a function of applied electric field for a 5000 Å thick $Alq_3$ device. The photocurrent was measured for 3.25 eV (380 nm) light incident through the Al electrode.

**Fig. 11.** Plot of the phase of the photocurrent as a function of applied electric field for a 5000 Å $Alq_3$ device.

**Fig. 12.** A semilog plot of ($J_d$ / F) vs $F^{1/2}$ for a 1000 Å ITO/$Alq_3$/LiF/Al device in forward bias.

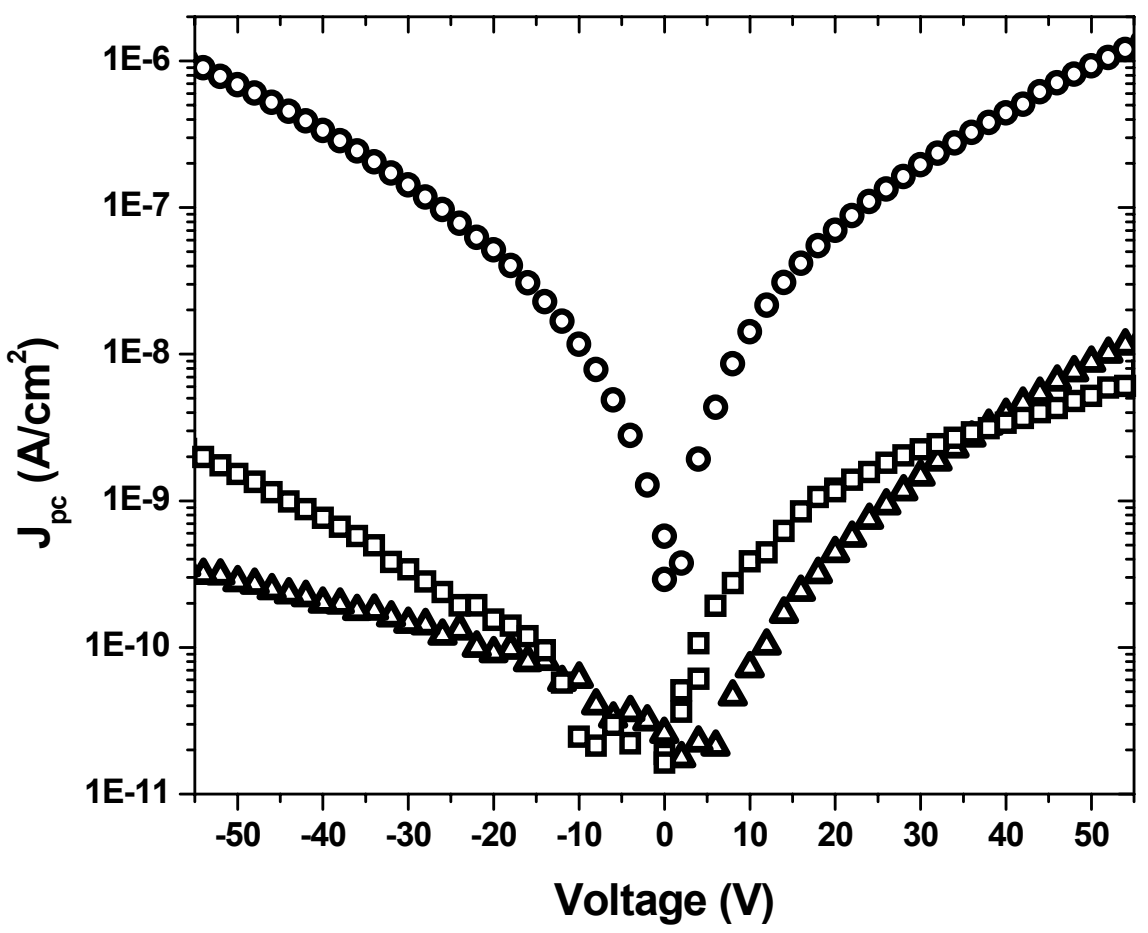

**Fig. 1.** Debdutta Ray et al.

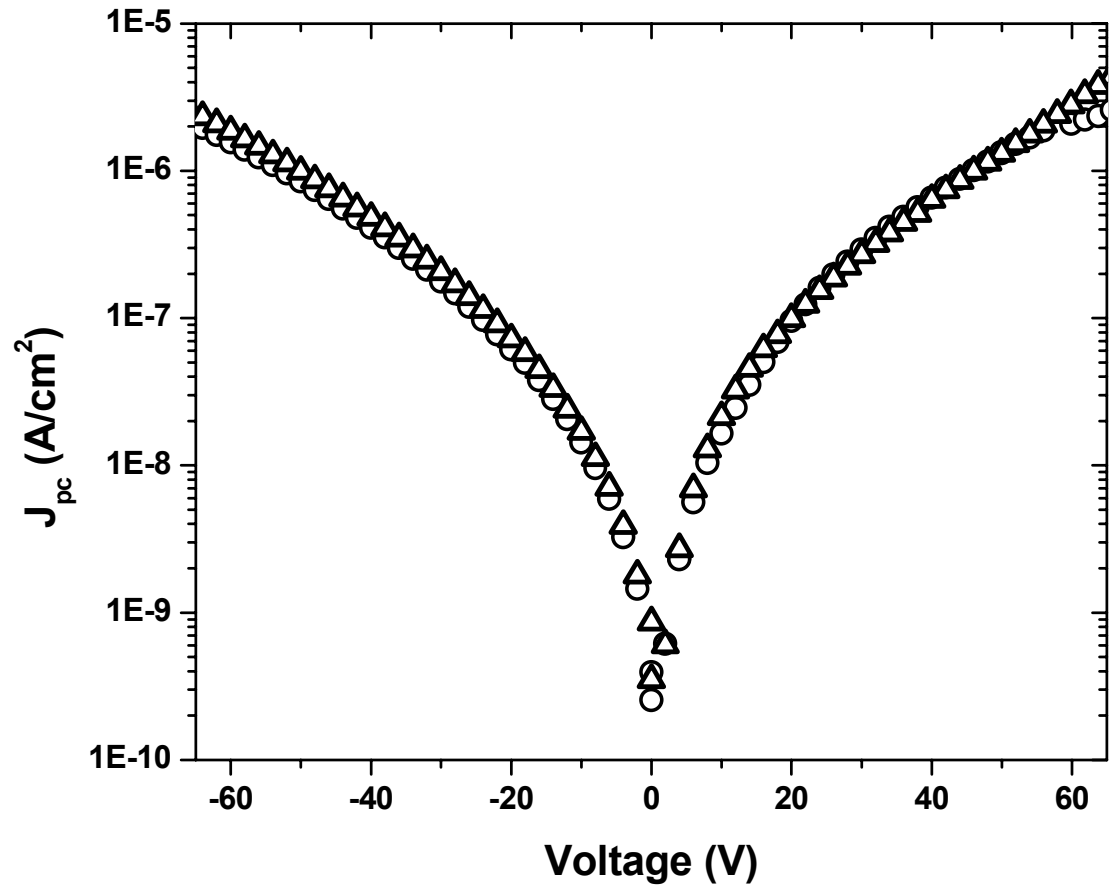

**Fig. 2.** Debdutta Ray et al.

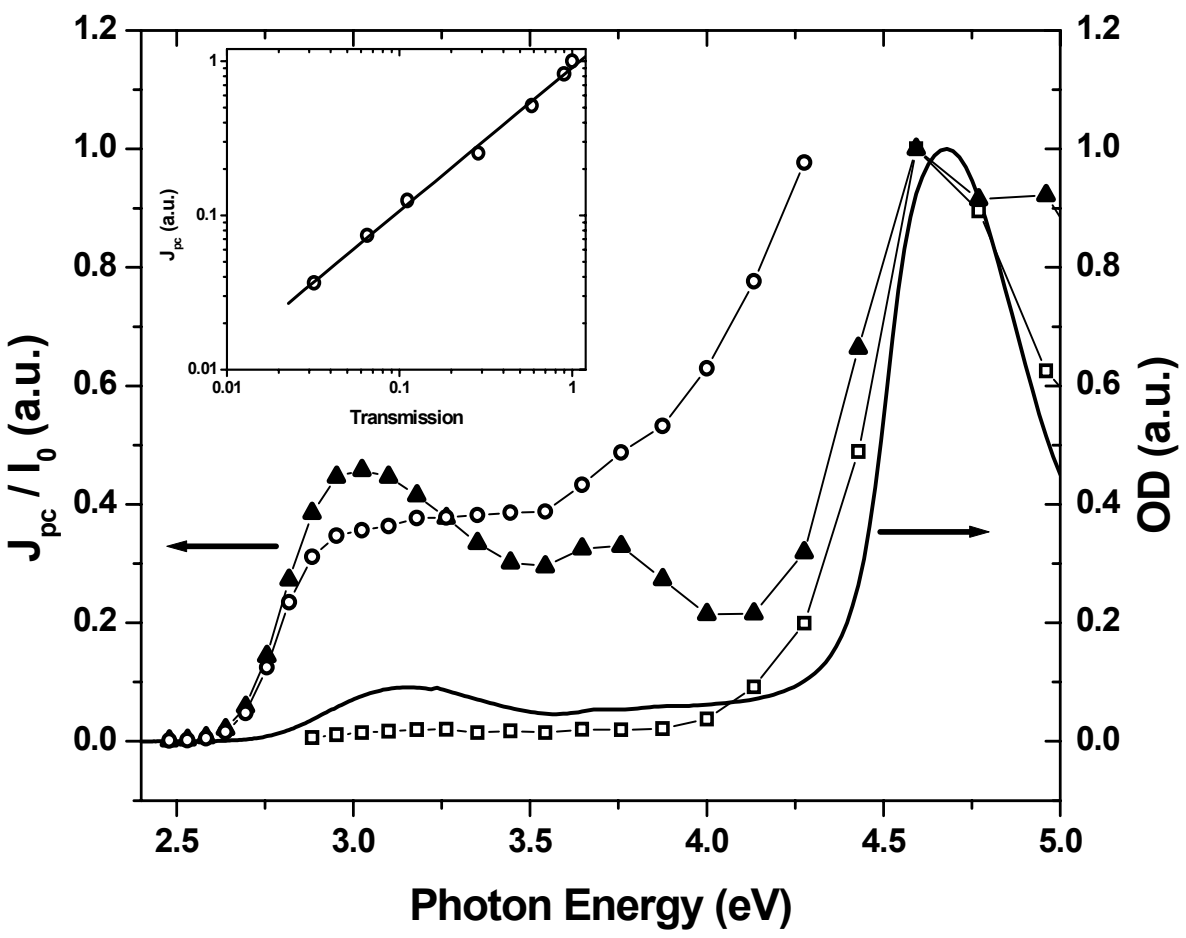

**Fig. 3.** Debdutta Ray et al.

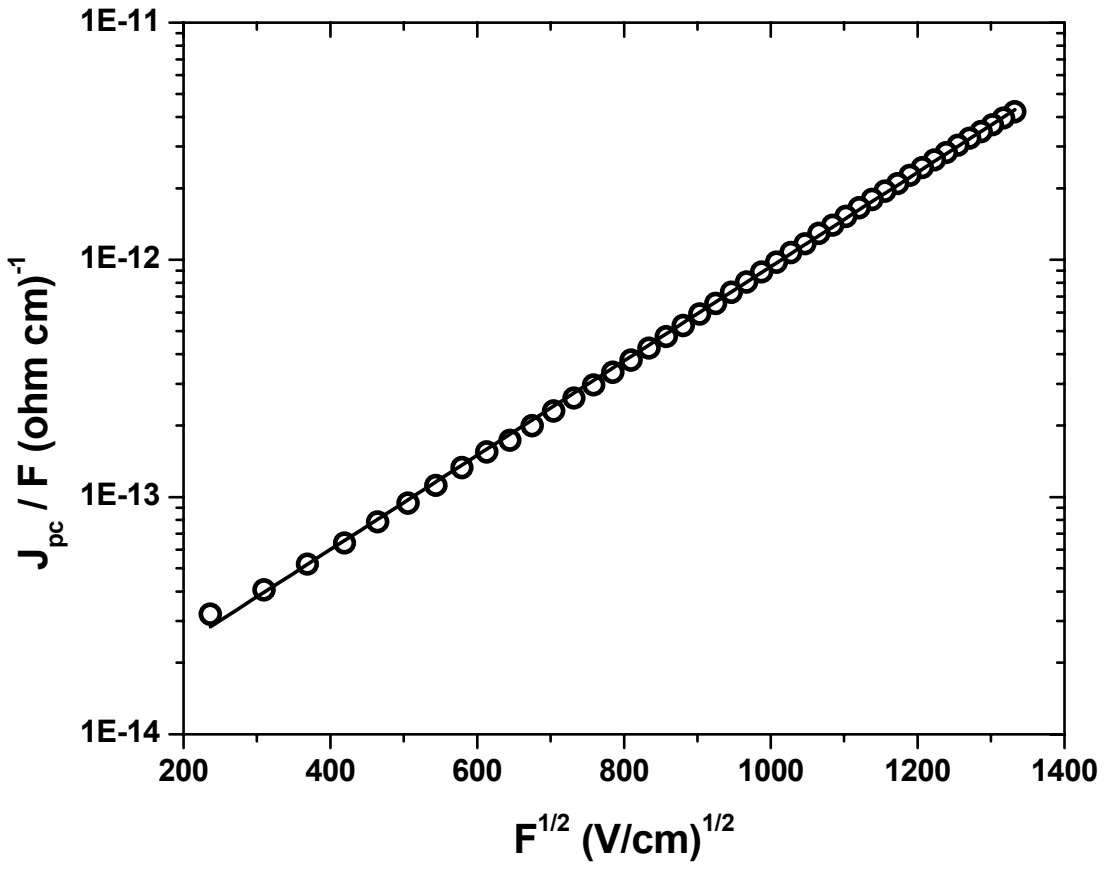

**Fig. 4**. Debdutta Ray et al.

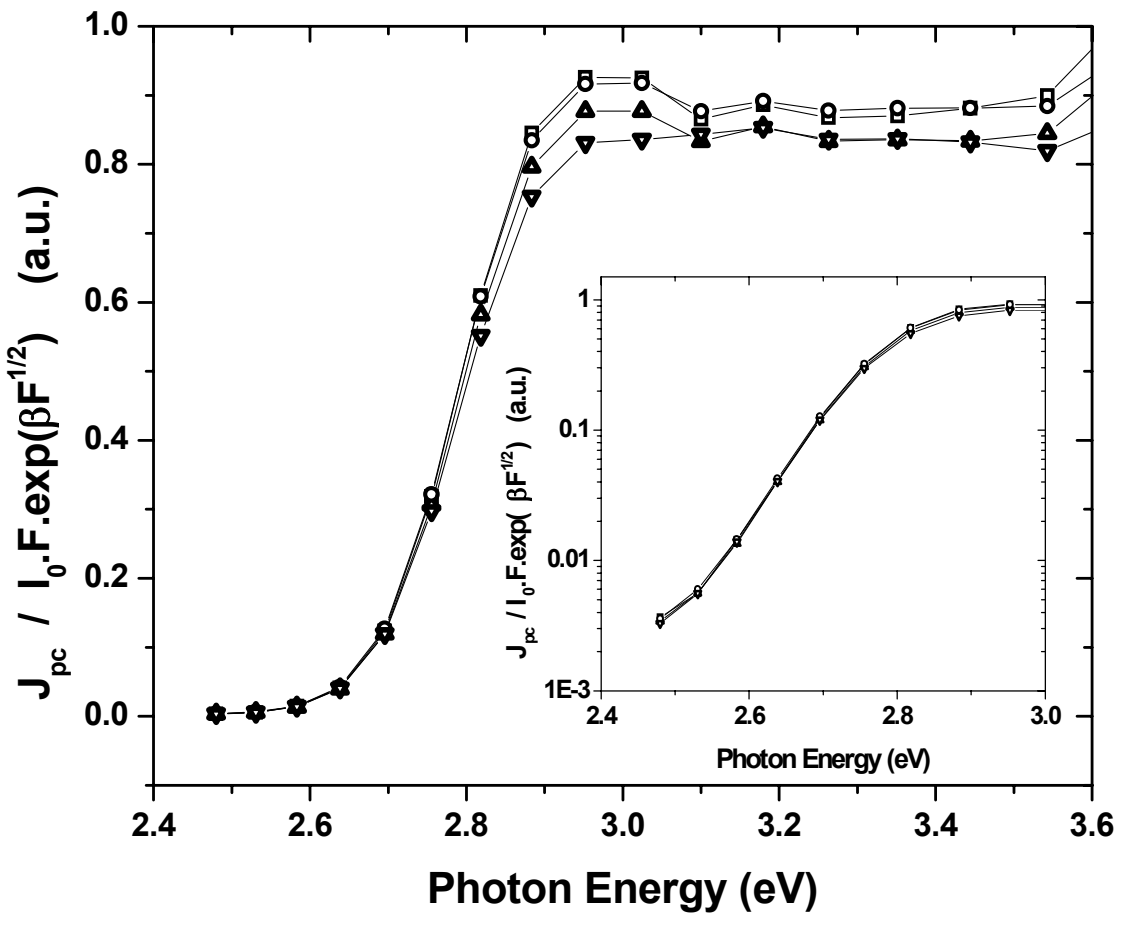

**Fig. 5.** Debdutta Ray et al.

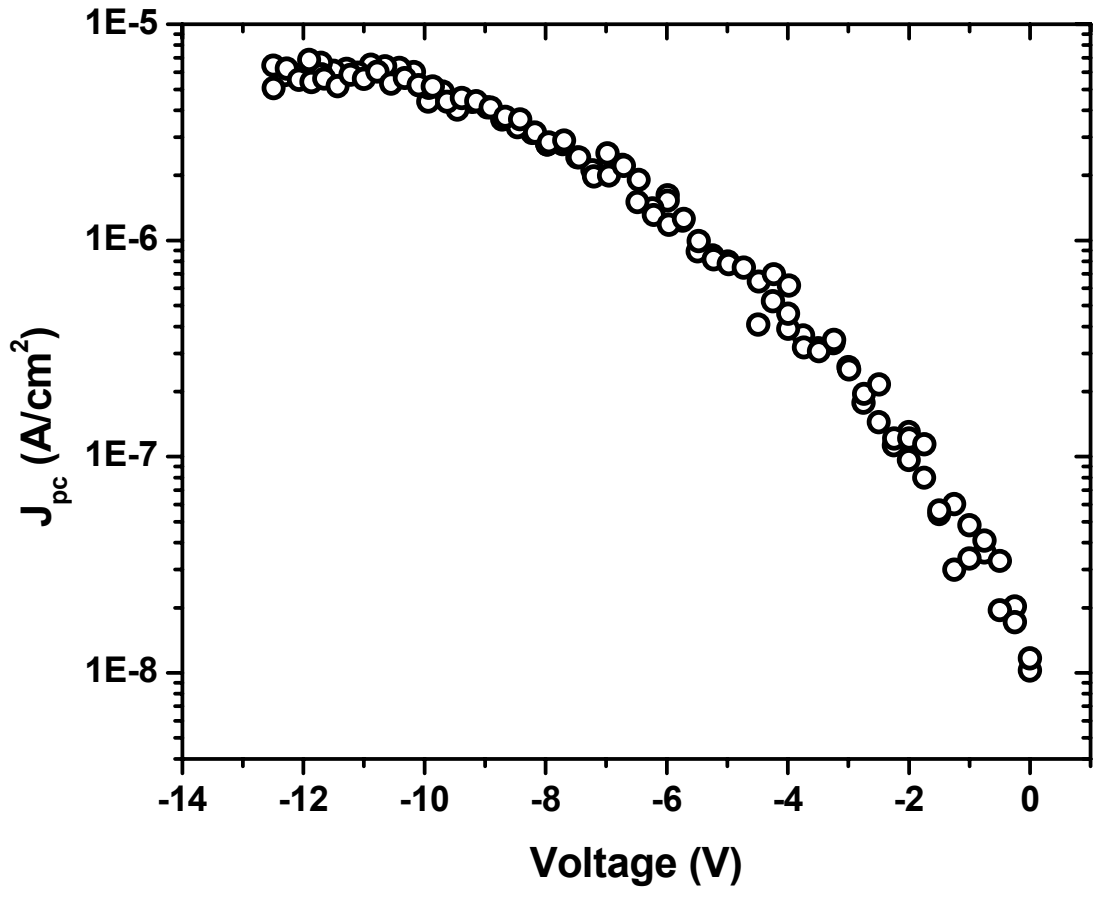

**Fig 6.** Debdutta Ray et al.

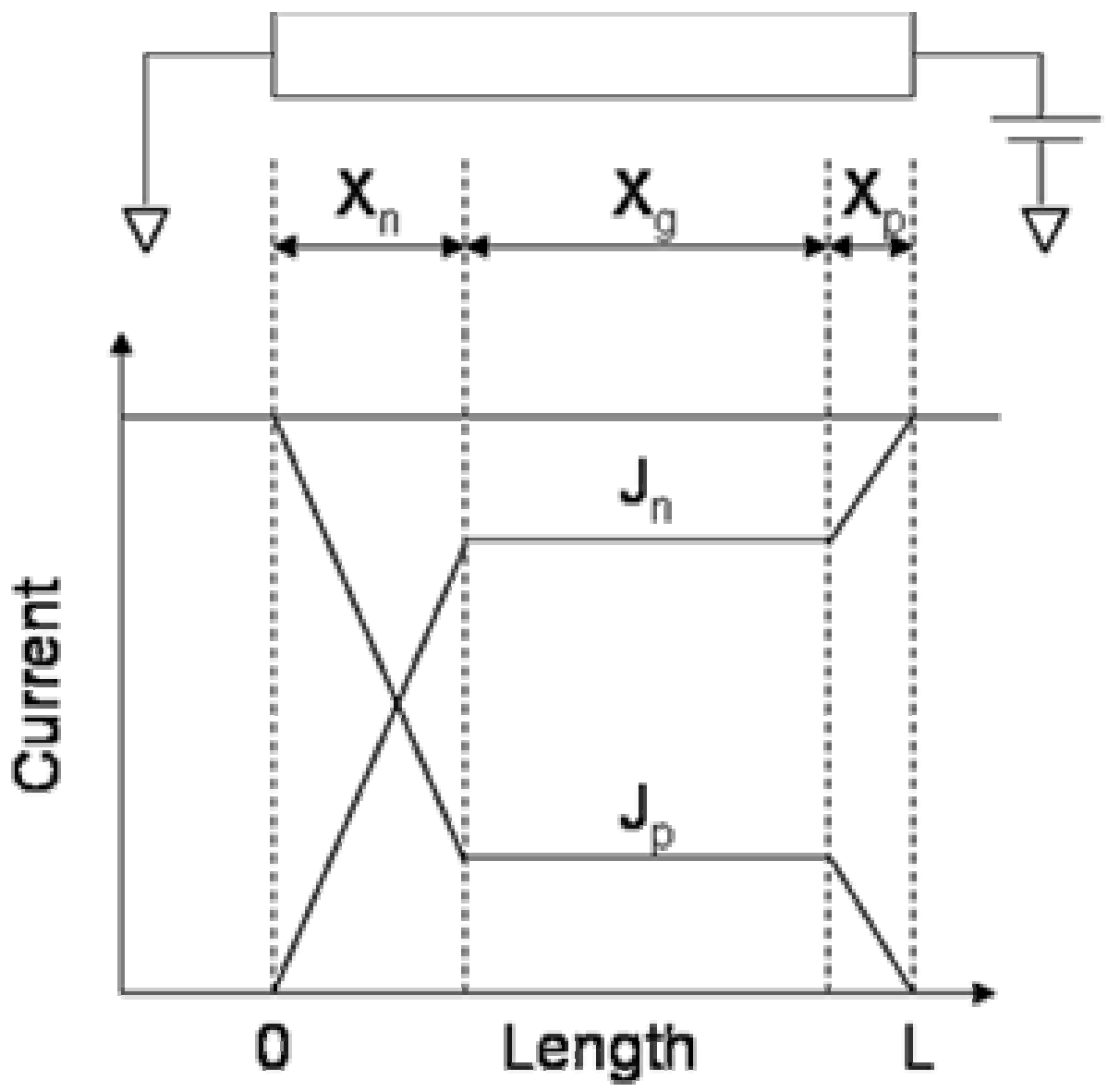

**Fig. 7.** Debdutta Ray et al.

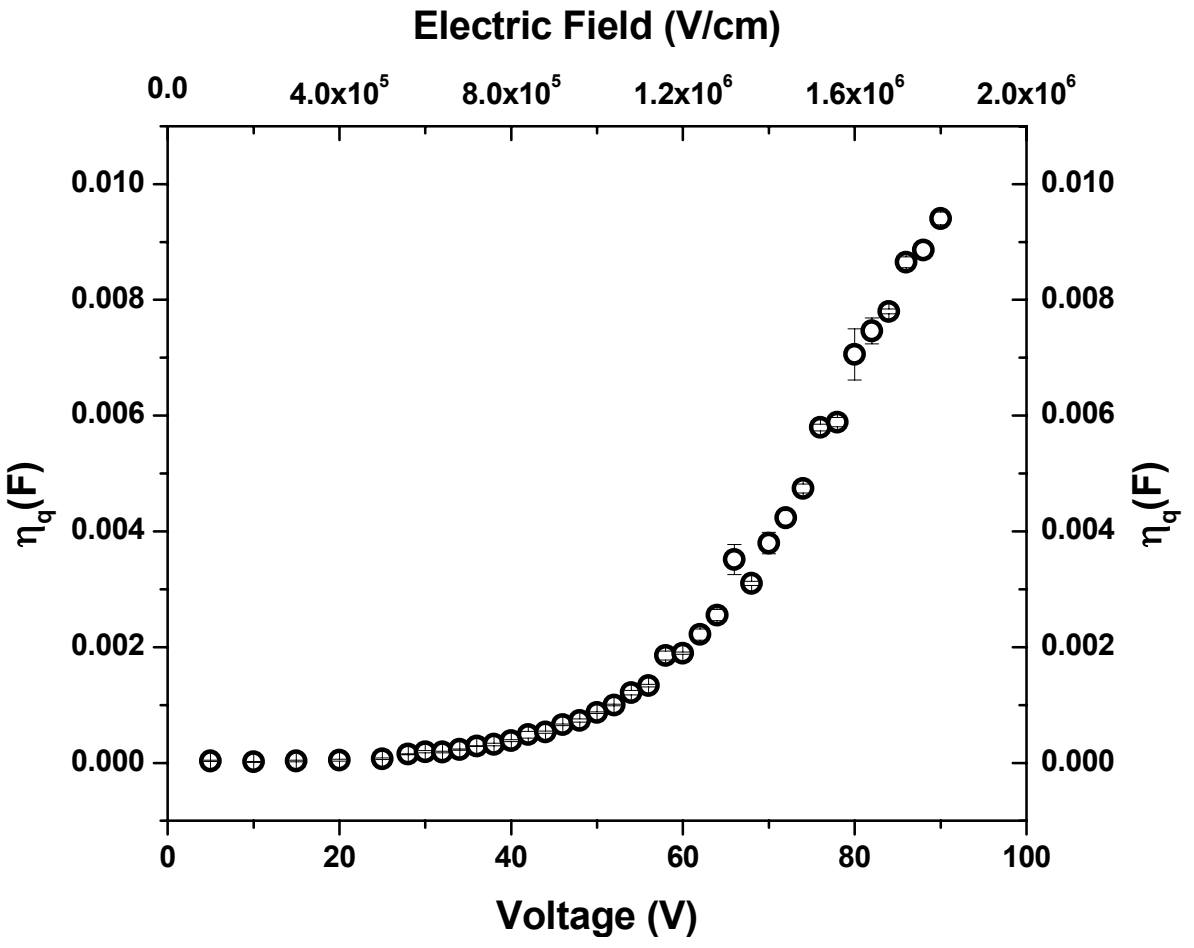

**Fig. 8.** Debdutta Ray et al.

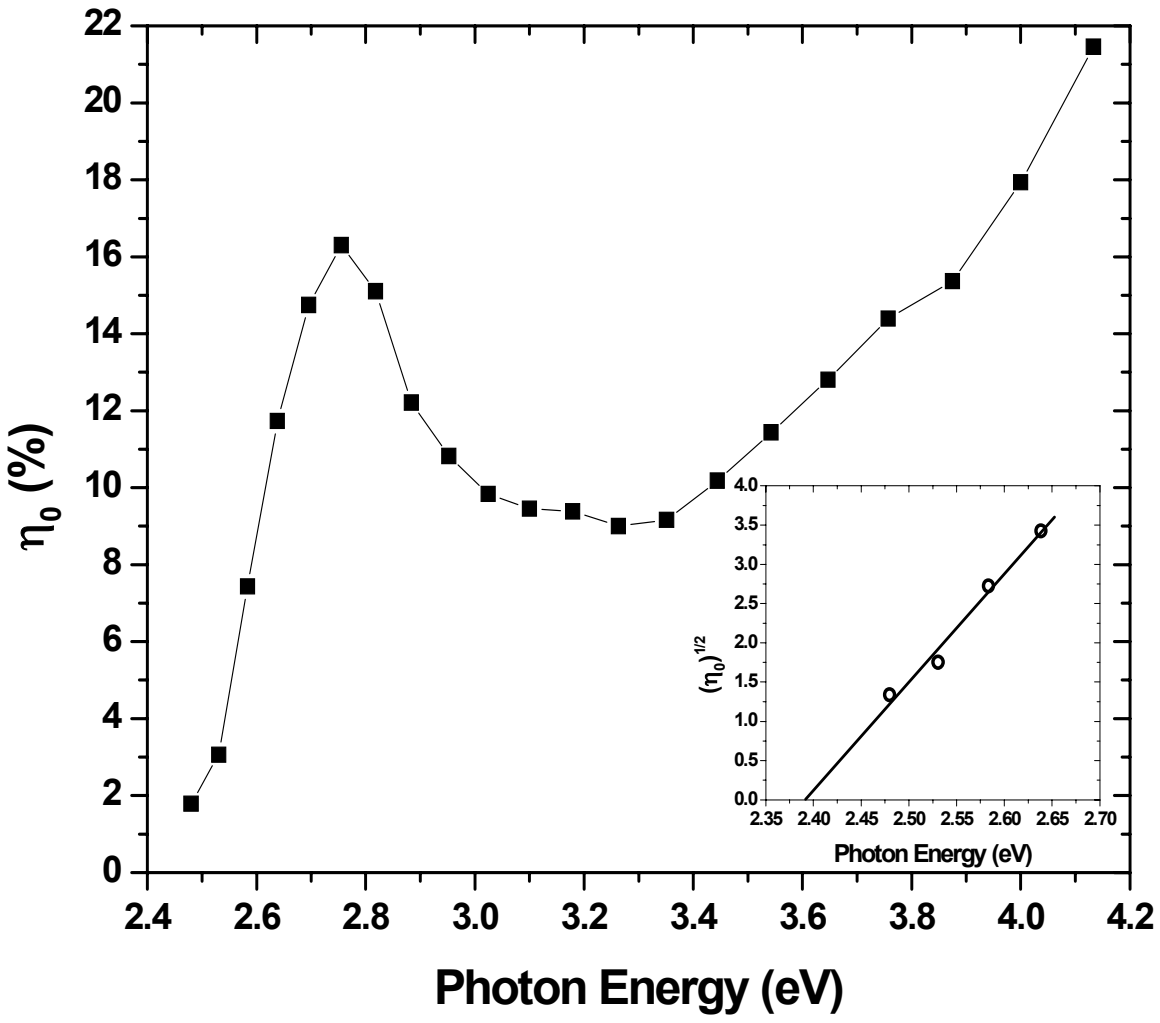

**Fig. 9.** Debdutta Ray et al.

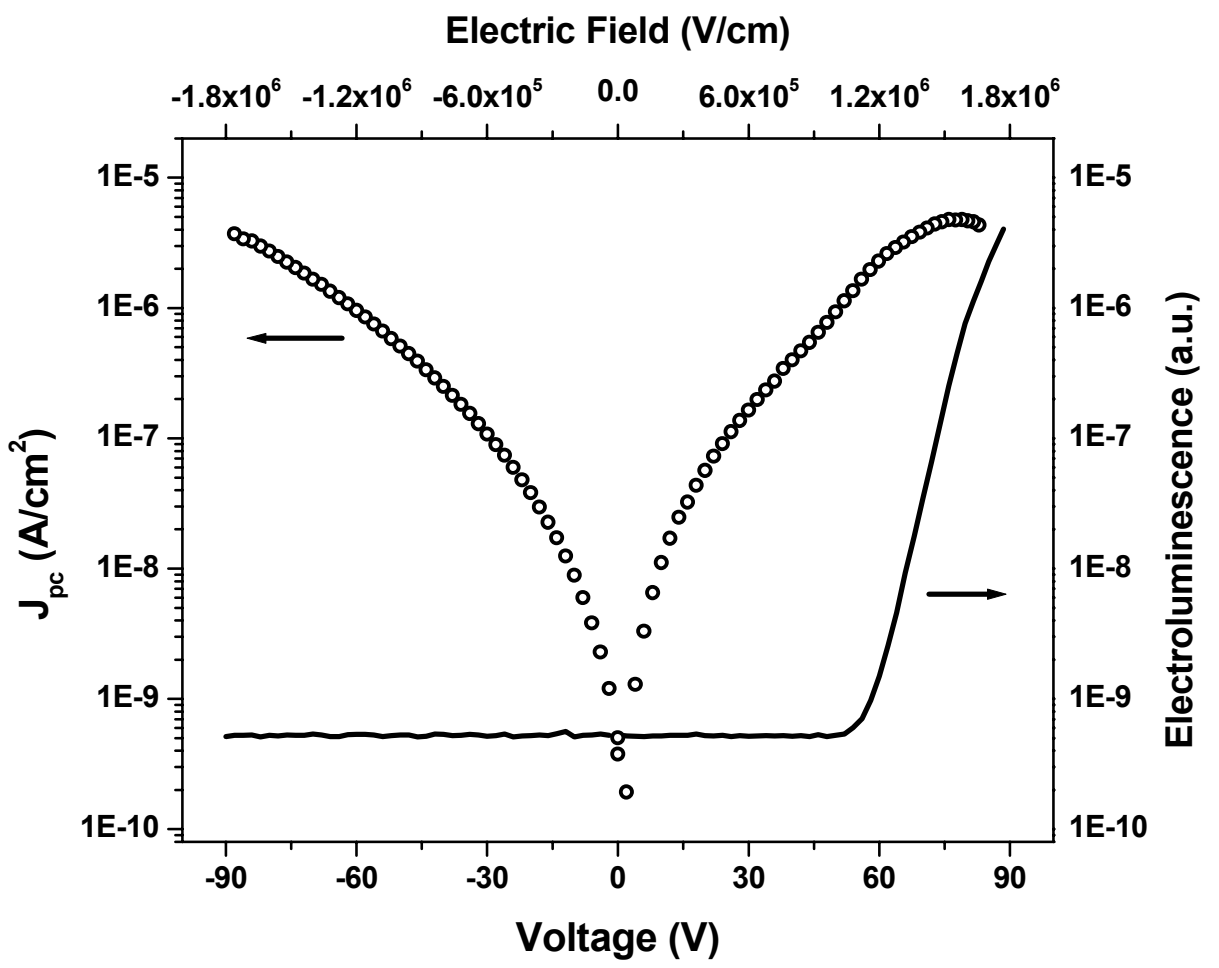

**Fig. 10.** Debdutta Ray et al.

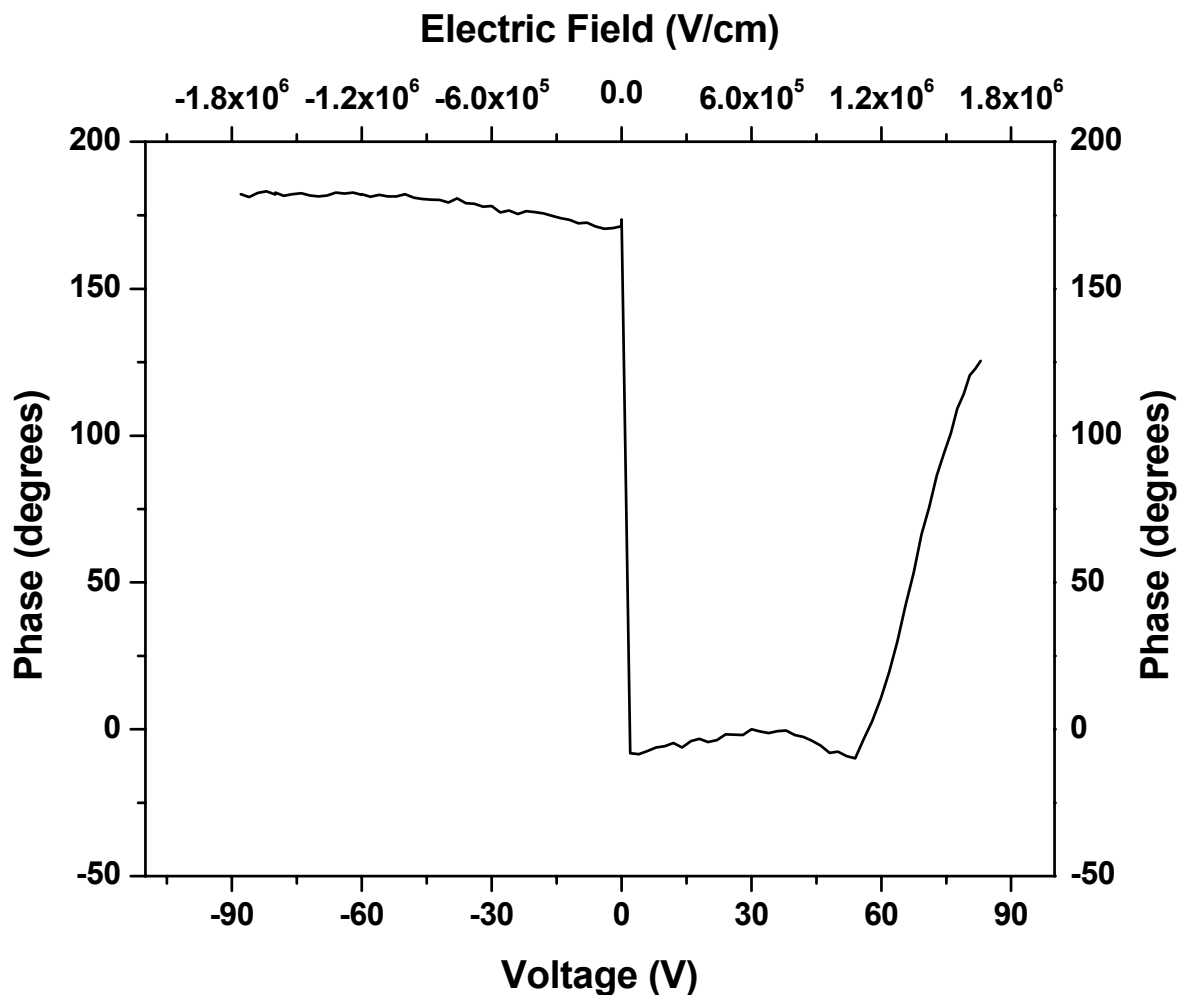

**Fig. 11.** Debdutta Ray et al.

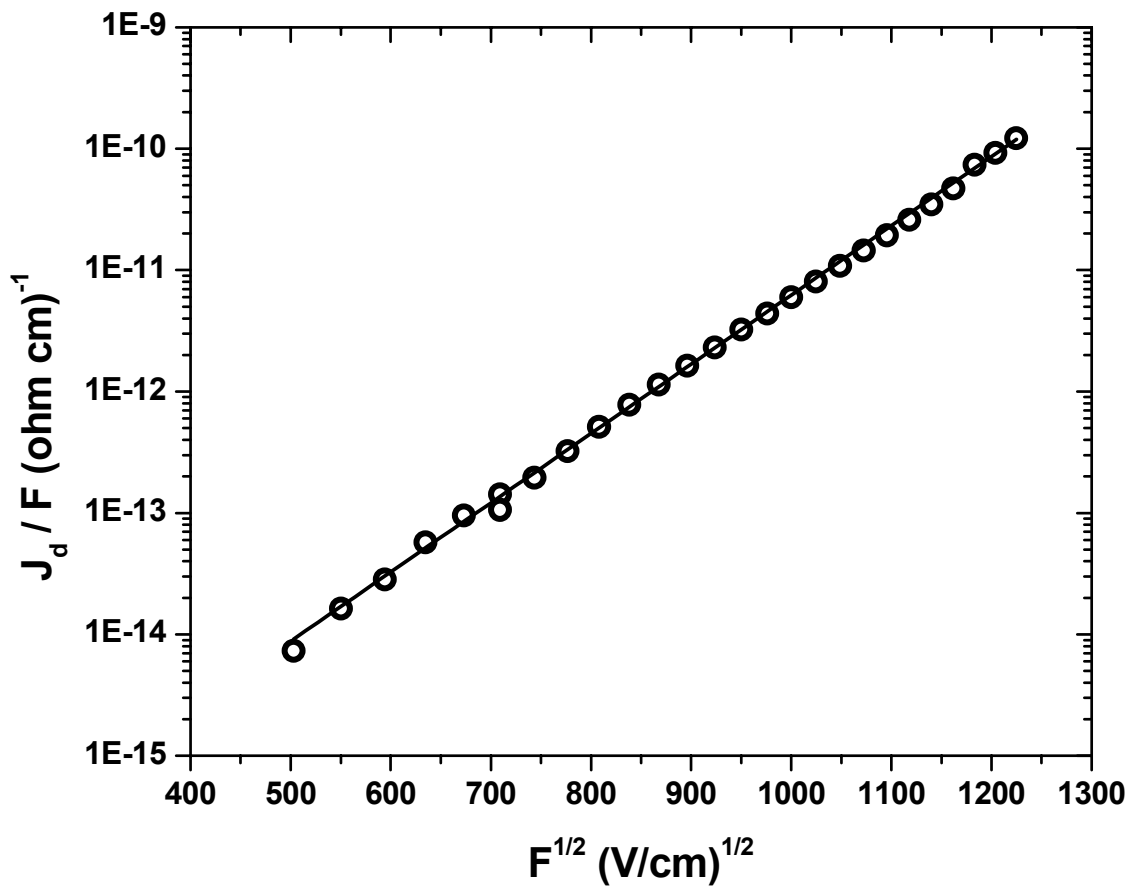

**Fig. 12.** Debdutta Ray et al.